# Defining mass transfer in a capillary wave micro-bioreactor

**Authors:**

Lasse Jannis Frey[a,c]

David Vorländer[a,c]

Detlev Rasch[a,c]

Sven Meinen[b,c]

Bernhard Müller[d]

Torsten Mayr[d]

Andreas Dietzel[b,c]

Jan-Hendrik Grosch[a,c]

Rainer Krull[a,c,]*

**Affiliation:**

[a] Institute of Biochemical Engineering, Technische Universität Braunschweig, Rebenring 56, 38106 Braunschweig, Germany

[b] Institute of Microtechnology, Technische Universität Braunschweig, Alte Salzdahlumer Str. 203, 38124 Braunschweig, Germany

[c] Center of Pharmaceutical Engineering (PVZ), Technische Universität Braunschweig, Franz-Liszt-Str. 35a, 38106 Braunschweig, Germany

[d] Institute of Analytical Chemistry and Food Chemistry, Graz University of Technologies, Stremayrgasse 9/II, 8010 Graz, Austria

* Corresponding author: Institute of Biochemical Engineering, Rebenring 56, 38106 Braunschweig, Germany

Email: r.krull@tu-braunschweig.de (R. Krull)

**Abstract**

For high-throughput cell culture and associated analytics, droplet-based cultivation systems open up the opportunities for parallelization and rapid data generation. In contrast to microfluidics with continuous flow, sessile droplet approaches enhance the flexibility for fluid manipulation with usually less operational effort. Generating biologically favorable conditions and promoting cell growth in a droplet, however, is particularly challenging due to mass transfer limitations, which has to be solved by implementing an effective mixing technique.

Here, capillary waves induced by vertical oscillation are used to mix inside a sessile droplet micro-bioreactor (MBR) system avoiding additional moving parts inside the fluid. Depending on the excitation frequency, different patterns are formed on the oscillating liquid surface, which are described by a model of a vibrated sessile droplet. Analyzing mixing times and oxygen transport into the liquid, a strong dependency of mass transfer on the oscillation parameters, especially the excitation frequency, is demonstrated. Oscillations at distinct capillary wave resonant frequencies lead to rapid homogenization with mixing times of 2 s and volumetric liquid-phase mass transfer coefficients of more than 340 $h^{-1}$. This shows that the mass transfer in a droplet MBR can be specifically controlled via capillary waves, what is subsequently demonstrated for cultivations of *Escherichia coli* BL21 cells. Therefore, the presented MBR in combination with vertical oscillation mixing for intensified mass transfer is a promising tool for highly parallel cultivation and data generation.

**Keywords**

Micro-bioreactor; mixing; $k_L a$; oscillation; capillary waves; droplets

**Highlights**

- Capillary waves induced by vertical oscillation are used to mix a sessile droplet micro-bioreactor
- Oscillations in resonance lead to specific mode patterns on liquid surface
- Mass transfer inside an oscillated liquid droplet is particularly affected by the excitation frequency

## 1. Introduction

Droplet-based cultivation systems have great potential to meet the overwhelming demand for experimental data in future research, especially for high-throughput biotechnology applications. Since each droplet can be seen as an independent reactor, numerous separate reactions can be performed simultaneously [1–4]. Enabling a massive increase in parallel analysis, droplet microfluidic technologies have developed into two distinct directions: The droplets are either a) generated in enclosed micro-channels and conveyed via a continuous fluid flow or, b) manipulated while being pinned to a planar solid surface [1,5]. The first operation mode, droplets in pressure driven flow (a), results in permanent movement of the separate droplets, i.e. reaction elements, what implies challenges especially for sensor integration and fluid control [3]. In case of pressure driven flows, rapid mixing purely by diffusion can only be achieved in extremely small droplets [6] or by Taylor vortices when the moving droplets remain in contact with the channel walls [7]. The second model (b) is characterized by sessile droplets, which are formed due to the surface tension of the fluid and interactions at the liquid-solid interphase. Depending mostly on the hydrophobicity of the solid and the resulting contact angle of the fluid, the form of the droplet can vary from rather flat puddles to spherical shapes [8]. These approaches are also known as *digital microfluidics* (DMF) [9,10]. Sessile droplets are generally larger (nL to µL range) compared to droplets in classical flow-microfluidics, where mostly pL to nL are handled. Hence, the generation of sessile droplets can be performed using regular pipettes or piezoelectric ejectors [1,2] resulting in an ease of operation. If smaller volumes are to be dosed, techniques from microfluidics can be adapted using T-junctions [11,12], flow-focusing devices [13,14] or microinjection [15]. Additionally, in DMF each droplet can be controlled individually without the need for networks of channels, pumps or valves, which separates them from classical flow-based microfluidics [10]. Due to the simpler and more compact design options, flexibility for application is increased [1,5,10]. Sessile droplet approaches benefit from small fluid volumes, a high capacity for parallelization and automation, lower sample consumption and straightforward integration with existing analytical techniques [1,16]. This technique therefore emerges as a platform capable of competing with microplates in terms of versatility and simplicity of operation [1].

The versatility of sessile droplet systems has been proven in various applications within a wide range of fields [10]. In chemical and enzymatic synthesis, reactions have been performed in droplets to study kinetics or evaluate new compounds [9,17,18]. Furthermore, the reduced sample volume is especially advantageous for analytical applications such as immunoassays [19–21], DNA-based applications, including sequencing, extraction or hybridization [22–27], or clinical diagnostics [28,29].

In addition, DMF has already been applied for living cells as well, including bacterial, yeasts and algae cells [30–32], mammalian cells [33,34] or even cell spheroids and 3D-cell culture [10,35–37].

Especially for the cultivation of cells but also for other applications, rapid mixing and homogeneous reaction conditions are prerequisite to avoid temperature or mass gradients and ensure sufficient substrate supply [38,39]. Here, the pressing matters are the prevention of cell sedimentation and the availability of nutrients for the cells, since diffusive transport is mostly insufficient and too slow [40]. Moreover, ensuring sufficient transport of oxygen to the liquid phase is considered as one of the key challenges for bioreactor design caused by its low solubility as well as the high demand in aerobic bioprocesses [41,42]. Consistent and reproducible results can only be achieved, if the mass transfer is sufficiently high, avoiding limitations hindering cell growth and product formation. This has to be achieved by applying an effective active mixing strategy [43,44].

Implementing active mixing inside a discrete sessile droplet, however, bears certain challenges due to the small fluid volumes, accompanied by the absence of turbulent flows, and the lack of a solid reactor surrounding [45,46]. Here, viscosity and capillary forces increasingly dominate the influence of gravitational and inertial forces as the according system dimension is reduced [3,47]. Furthermore, the implementation of mixing parts inside a droplet requires extremely challenging 3D microfabrication. In order to effectively introduce advective transport into sessile droplets avoiding movable parts in the reaction fluid, various techniques have been developed and reported. Droplets can be actuated through the application of electrical potential by implementing a pair of electrodes, either in planar configuration or with a top hanging wire, which is commonly termed electrowetting on dielectric [5,48,49]. Additionally, piezoelectric transducers are used to excite the fluid in resonance at frequencies in the kilohertz range or higher [50]. Acoustic streaming and surface acoustic waves are another commonly used techniques for mixing small fluid volumes [45,51–53]. For all these techniques, energy is transmitted via oscillations exciting the phase boundary to resonate. The technical effort as well as the operation of high frequency systems, however, poses significant drawbacks. Due to the increased energy input, temperatures in small fluid volumes tend to increase [50]. Additionally, cells can get damaged or disrupted by high frequency or even ultrasound vibrations. In addition, the fabrication requires special care for shielding the electrical devices.

Circumventing the issues of high frequency actuation, another valuable approach for mixing sessile droplets is based on the application of low frequency vertical oscillation, causing the gas/liquid interphase to oscillate in resonance. If a fluid is vertically vibrated in regimes where inertial forces and surface tension compete, the drop itself oscillates [54], leading to a stationary surface wave on the fluid, which subsequently leads to bulk mixing [55]. These stationary surface waves on the fluid with appropriate wavelengths are termed *capillary waves* [56]. The decisive factor for the energy dissipation in an oscillating system is the excitation frequency. In general, a higher excitation frequency leads to a higher wavenumber or higher number of so called nodes, which are locations of minimum amplitude. The wavelength decreases correspondingly [54,57]. Noblin et al. [58] examined the effects of vertical vibration on a sessile droplet using a pure sinusoidal excitation. The postulated model can be adapted to calculate resonance frequencies ($f_n$) for one-dimensional waves for a fluid of height $h$ in the capillary-gravitational region [56,58], shown in Eq. (1). The angular frequency is here substituted with the expression $2 \cdot \pi \cdot f_n$ and rearranged for $f_n$.

$$f_n = \sqrt{\left(\frac{g \cdot n}{4 \cdot \pi \cdot L} + \frac{\gamma}{\rho} \cdot \frac{\pi \cdot n^3}{4 \cdot L^3}\right) \cdot \tanh\left(\frac{\pi \cdot n \cdot h}{L}\right)} \qquad (1)$$

Here, $g$ is the gravitational force, $n$ half of the wavelength on the droplet surface, $L$ the length of the surface profile and $\gamma$ and $\rho$ the surface tension and the density of the fluid, respectively. Oscillation in resonance leads to unique modes, which is defined by the number and position of the nodes [54,57,58].

The resulting oscillation mode patterns on a vibrated droplet, however, cannot be sufficiently accurately described by a one-dimensional wave due to the asymmetry of the oscillations. The use of a resonance mode numbering scheme with a single value $n$, as performed by Noblin et al. [58], Temperton and Sharp [59], as well as Kim and Lim [60] cannot completely account for the three-dimensional nature of the oscillating interphase, which is not axisymmetric to the cross sectional area. To more precisely describe the oscillation modes, the second dimension has to be included, resulting in a second oscillation axis. Mile et al. [55] described such non-axisymmetric oscillations of spherical drops with two oscillation axis. These involve both a degree and an order number, which cannot be directly mapped to values of the resonance modes $n$. A similar concept was postulated by Chiba et al. [61] for sessile droplets attached to a conical base, having tilted sidewalls constraining the liquid. Here, a meridian directional mode number as well as a zenith angle directional mode number, meaning two oscillation directions, are used to describe the modes of asymmetric oscillations. An even more comprehensive study to describe oscillations of spherical sessile droplets is reported by Bostwick and Steen [54,57]. The authors report modal shapes identified by two variables, a polar and azimuthal (circular) wavenumber. With the resulting nodal lines and nodal circles two-dimensional mode patterns can be described.

Practical implementations of low frequency vertical oscillation has already proven to mix abiotic sessile droplet systems sufficiently [59,60,62,63]. In previous work of this working group the vertical electrodynamic oscillation technique was initially applied on a micro-well based bioreactor resulting in an outstanding mixing performance [44]. Subsequently, it was used in combination with a micro-fabricated *capillary wave micro-bioreactor* (cwMBR) with a fluid volume of 7.3 µL, where the liquid droplet is restrained in a trough ensuring a defined positioning of the fluid as well as reproducible droplet shapes. The resulting fluid velocities as well as typical flow patterns occurring in resonance were determined using stereo micro particle image velocimetry [64]. Here, the one-dimensional model was shown to adequately predict the resonance frequencies. Additionally, the number of flow vortices was shown to increase for higher resonances.

In this research, the excited capillary waves and the resulting mode patterns on the liquid surface will be described using a two-dimensional model of a sessile droplet. The cwMBR with the according oscillation mixing will be characterized for its mixing performance and the oxygen mass transfer into liquid phase, being key parameters for biological cultivation systems. On this basis, it will be clarified whether mass transfer can be tailored by vertical oscillation and the resulting capillary waves. Subsequently, the impact of oscillation settings on the cultivation performance of *Escherichia coli* BL21 will be shown using online monitoring of biomass concentration observed via the scattered light intensity and dissolved oxygen (DO).

## 2. Materials and methods

*2.1. Capillary wave micro-bioreactor and vertical oscillation mixing setup*

The *capillary wave micro-bioreactor* (cwMBR) was manufactured in Foturan® glass (Schott, Mainz, Germany) using femtosecond (fs) Laser Direct Writing (LDW) as previously reported by Meinen et al. [64]. By special thermal post-treatment, a very low surface roughness is achieved leading to enhanced biological and chemical inertness as well as to high transparency, which is advantageous for the integration of optical sensors. The cwMBR represents a sessile droplet concept, however the fluid droplet is not placed on a flat surface but into a trough-shaped reactor cavity geometrically holding a volume of 7.3 µL (**Fig. 1**). The cwMBR shape leads to a precise positioning of the fluid as well as formation of a defined droplet curvature and it hinders any horizontal movement. Since it is open from the top, it can be easily filled from above and is additionally accessible via a micro-channel positioned at the reactor bottom.

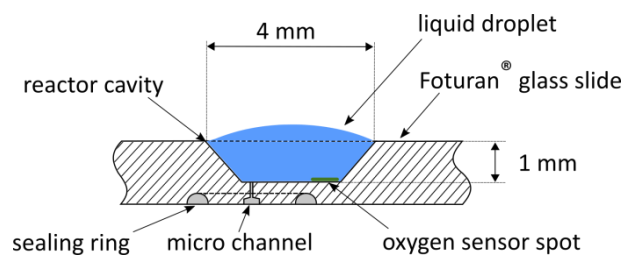

**Figure 1**: Schematic illustration of the cwMBR in cross sectional view. The reactor cavity with a circular cross-section of 4 mm at the top, manufactured into a glass slide, is filled with approximately 8 µL fluid, leading to a convexly formed droplet cap. The cwMBR can be filled from the top using a pipet or via a microfluidic channel at the reactor bottom. The oxygen sensor spot is read out non-invasively by an optical fiber from beneath the reactor slide.

The cwMBR was fixed in a 3D-printed mounting (Agilista 3200W, Keyence, Osaka, Japan), made of semitransparent AR-M2 polymer, allowing for reliable and precise integration of optical fibers for sensor read-out (**Fig. 2**). The reactor chip is embedded into a base element and clamped with a lid, which has four water troughs in the walls around the reactor cavity to increase humidity in the headspace to minimize evaporation of the reaction liquid.

Biomass was monitored using scattered light signal. Therefore, a 400 µm optical fiber (FT400UMT, Thorlabs, Dachau, Germany) coupled to an LED with 650 nm wavelength was positioned at an angle of 45° to the reactor bottom. Light scattering was then measured via a second optical fiber (90° to the excitation fiber) using a miniature spectrometer (USB2000+, Ocean Optics, Ostfildern, Germany). In this configuration the light reflectance of the gas/liquid interphase was reduced and cell growth resulted in a linear signal, as previously reported in the literature [65–67].

To online monitor the dissolved oxygen (DO) concentration, a sensor layer containing platinum(II) meso-tetra(4-fluorophenyl)tetrabenzoporphyrin (PtTPTBP) was inkjet-printed on the flat reactor bottom [68–70] and read out via a vertical optical fiber connected to a commercial miniaturized USB phase fluorimeter (FireStingO$_2$, PyroScience, Aachen, Germany). The signals were analyzed using the associated software (OxygenLogger, PyroScience, Aachen, Germany). Calibration was performed at 37 °C with air saturated water or, using gaseous nitrogen, oxygen stripped deionized (DI) water. To minimize evaporation of the reaction fluid, the inlet air was filtered and pre-humidified (Midisart 2000, Sartorius Stedim, Göttingen, Germany) before leading into the reactor headspace. Gassing of the cwMBR is solely performed via the headspace and mass transfer via the liquid surface.

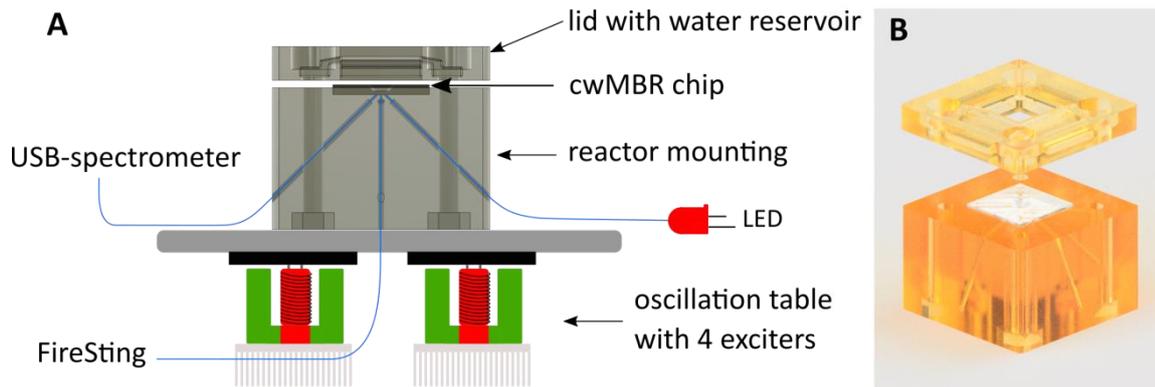

**Figure 2:** Schematic illustration of the cwMBR setup: (A) Side view of reactor mounting with optical fiber side-in module and sensor assembly of optical measurements. (B) Perspective view on rendered cwMBR mounting with base element, cwMBR and lid, having four water troughs in the walls. All parts are clamped together using four screws.

Mixing was performed via vertical oscillation as reported by Frey et al. [44] and Meinen et al. [64]. The reactor mounting (**Fig. 2**) was therefore fixed to a 20 x 20 cm polyvinyl chloride (PVC) panel, which was oscillated by the electromagnetic exciters situated beneath. The input signal for the exciters was created digitally (SoundcardScope, Zeitnitz, Essen, Germany) controlling frequency and the normalized amplitude. The latter is later being referred to as *excitation strength*.

To control the ambient conditions, the cwMBR together with the mixing setup was integrated into an incubation chamber. Using a Pt100 element connected to a heating thermostat (Eco E4, LAUDA Dr. R. Wobser, Lauda Königshofen, Germany) temperature was measured and regulated via a heat exchanger (Hydro Series H55, Corsair Components, Fremont, USA). Humidity in the chamber was kept at least at 93 % saturation using an ultrasonic humidifier (DH-24B, Conrad Electronic, Germany).

*2.2. Analysis of mixing time*

The mixing time ($t_M$) was determined at room temperature using a colorimetric method combined with image analysis. The cwMBR was therefore filled with 8 μL of DI-water (Astacus2 LS μS-control, membraPure, Hennigsdorf, Germany) and subsequently 200 nL of an ink-solution (4001, Pelikan, Hanover, Germany) were manually injected at the reactor bottom using a syringe (Trajan Scientific, Victoria, Australia). Homogenization was then recorded using a board camera (DFM 72BUC02-ML, The Imaging Source Europe, Bremen, Germany), which was positioned onto the reactor mount directly above the reactor cavity, leading to videos without relative movement of the reactor. Due to the compact size and low weight, the vertical oscillation was not affected by the camera. To ensure sufficient illumination allowing for proper color detection, three white LEDs where integrated into the reactor mount, resulting in a brightfield image. An exemplary record of a mixing time experiment is shown in **Fig. 3**.

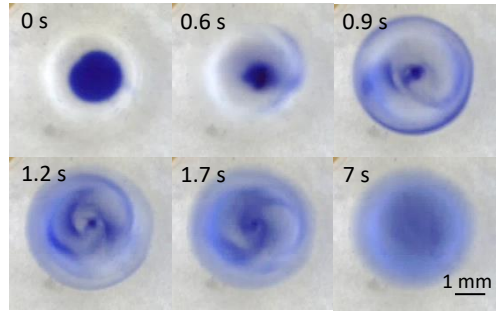

**Figure 3**: Series of pictures in top view of an exemplary colorimetric mixing experiment with 200 nL of ink solution placed at the bottom center of the reactor (top left). After starting the vertical oscillation, the fluid is homogenized within 7 s.

Using the software IC capture 2.4 (The Imaging Source Europe, Bremen, Germany), videos were captured at a framerate of 25 fps and subsequently analyzed using an in-house Python program, which makes use of the openCV library for image analysis. The procedure analyzing the greyscales for each pixel to determine $t_M$ is performed analogous to Frey et al. [44]. $t_M$ was then defined as the time difference between the initial frame and the frame at corresponding time, where 95 % homogeneity is reached [71,72]. All measurements were performed in triplicates between frequencies of 40 and 200 or 400 Hz respectively and excitation strength was varied between 10 and 30 %. To analyze the dependence of $t_M$ on the viscosity of the fluid, DI-water was replaced with glycerol solutions between 10 and 80 % (v/v).

*2.3. Analysis of volumetric mass transfer coefficient $k_La$*

The volumetric mass transfer coefficient $k_La$ of the cwMBR using vertical oscillating mixing was determined using the dynamic gassing-out method [38,40]. Excitation frequency and signal strength were varied between 0 and 360 Hz in steps of 10 Hz and between 5 and 20 %, respectively. Therefore, the cwMBR was filled with 8 µL dH$_2$O and the headspace was gassed with nitrogen with a flow rate of 1 L min$^{-1}$ until a constant signal was reached, which was recorded using the software Oxygen Logger (PyroScience, Aachen, Germany). Subsequently, gassing was switched to ambient air until the signal was constant again. The $k_La$ was calculated for a given time frame $t$ by Eq. (2):

$$k_L a = \frac{1}{t} \cdot \ln \frac{c^*_{O_2} - c_{0,O_2}}{c^*_{O_2} - c_{O_2}} \quad (2)$$

Here $c^*_{O_2}$ is the dissolved oxygen saturation concentration, $c_{0,O_2}$ is the dissolved oxygen concentration at $t$ = 0 h and $c_{O_2}$ is the measured dissolved oxygen concentration [40,41].

*2.4 Cultivating of Escherichia coli BL21 (DE3) pMGBm41 in the capillary wave micro-bioreactor*

Cultivations of *Escherichia coli* BL21 (DE3) with the plasmid pMGBm41 were performed using Lysogeny broth (LB) medium containing 10 g L$^{-1}$ tryptone, 5 g L$^{-1}$ yeast extract and 5 g L$^{-1}$ sodium chloride [73]. All chemicals were purchased from Sigma Aldrich (Steinheim, Germany). Inocula were grown overnight in 100 mL shaking flasks (Schott, Mainz, Germany), containing four baffles and filled with 10 % (v/v), at 37 °C with a shaking speed of 120 min$^{-1}$ and a shaking diameter of 5 cm. A second pre-culture was inoculated using the over-night culture with an optical density (OD) of 0.1, which was incubated until cells reached the exponential growth phase. Subsequently, 10 mL of fresh LB-medium were inoculated to a starting OD of 0.1, of which 8 µL were transferred into the cwMBR. During cultivation ambient condition in the reactor chamber were kept constant at 37 °C and 93 % relative humidity. To minimize evaporation of the cultivation broth, water troughs in the cwMBR mounting lid were filled with sterile DI-water and the headspace was covered with a silicone foil. Oscillation parameters were varied according to the mass transfer requirements (frequencies 70, 110 and 150 Hz, 5 and 20 % excitation strength). DO was measured online every 60 s using the previously described sensor spot. Biomass was monitored via the scattered light intensity, where the signals were integrated over 5 s and 12 measurements were averaged, leading to a sampling rate of 1 min$^{-1}$.

## 3. Results and discussion

*3.1. Capillary wave modes in resonance on liquid surface*

The cwMBR (as shown in **Fig. 1**) constrains the liquid and supports the formation of a curved liquid surface resulting in a large interface between gas and liquid that exhibits a slightly convex form, if filled with more than 7.3 µL. Whereas for most of the reported sessile droplet approaches the liquid droplets are placed on a flat and free surface [1,5,10], the cwMBR precisely locates the liquid at a defined position. The sharp edge of the cwMBR effectively hinders the liquid from spilling or flattening out and the three-phase contact line is pinned on the upper edge of the cavity. The planar reactor bottom in combination with the highly transparent Foturan® glass facilitates the integration of optical sensors.

As previously described in Frey et al. [44] and Meinen et al. [64], vertical oscillation is used for homogenization. The cwMBR has to be vibrated within specific frequencies, in order to provoke an oscillating interface leading to resonance at which typical vortices patterns can be observed [64]. Using Eq. (1), which is adapted from Noblin et al. [58], the resonance frequencies ($f_n$) of the first 5 resonance modes (*n*) can be calculated (**Tab. 1**).

**Table 1:** Resonance frequencies of the first 5 resonance modes of the cwMBR calculated using Eq. (1).

| Resonance mode $n$ | Frequency $f_n$ [Hz] |
|:---:|:---:|
| 1 | 27.4 |
| 2 | 82.0 |
| 3 | 154.4 |
| 4 | 238.6 |
| 5 | 333.3 |

To calculate $f_n$, the acceleration of gravity $g = 9.81$ m s$^{-1}$ as well as the density and surface tension of water at room temperature were used, i.e. $\rho = 998$ kg m$^{-3}$ and $\gamma = 0.0728$ N m$^{-1}$. Additionally, for a filling volume of 8 μL the mean height of the droplet was estimated as the depth of the cavity [58], resulting in $h = 1$ mm and the profile length $L$, where the oscillation occurs, equals to 4.02 mm. Slight deviations to the resonance frequencies calculated in Meinen et al. [64] occur from a larger filling volume applied here.

If excited with frequencies close to the resonance frequencies shown in **Tab. 1**, capillary waves are formed on the liquid surface, as shown in **Fig. 4**. Here, excitation frequencies were evaluated for the strongest resonance in steps of 10 Hz, leading to slight deviation from the calculated values.

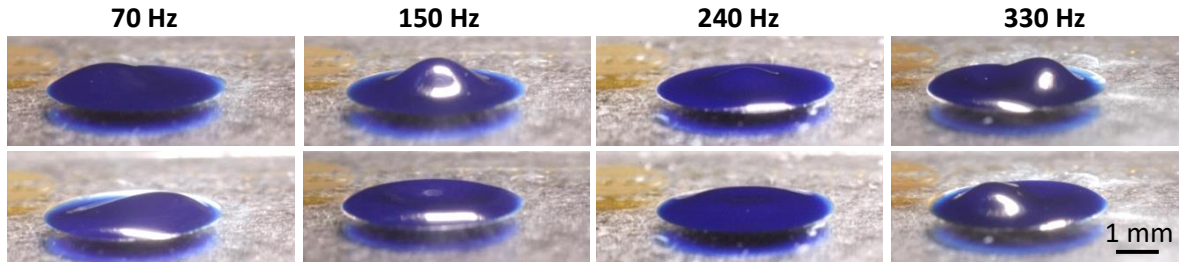

**Figure 4:** Perspective view on the cwMBR filled with dyed water excited at its resonance frequencies. Different oscillation patterns are formed on the liquid surface, whereby the wavenumber increases for higher excitation frequency. Two characteristic time points are shown, where the amplitude of the oscillating liquid interface is the highest. Images were taken with a single-lens reflex camera (EOS 60d, Canon, Tokyo, Japan) connected to a micro-Nikkor objective (Nikon, Tokyo, Japan) with a focal length of 55 mm and a triggered ultrashort time flash with an exposure time of $1 * 10^{-5}$ s.

Under the influence of competing surface tension and inertial forces, different oscillation modes are formed depending on the excitation frequency. Here, two characteristic time points are shown, where the oscillation amplitude of the liquid surface is the highest. These different patterns on the liquid surfaces are generally referred to as modes having characteristic nodes, which are transient regions between nearby up-down moving surfaces [60]. As described by Noblin et al. [58], the number of nodes increases for higher frequency. This holds true for the shown oscillations, where the wavenumber increases for higher frequencies. In Meinen et al. [64] it was shown, that the resonance frequencies can be predicted using the one-dimensional model with adequate precision. To precisely describe the excited mode patterns on the liquid surface, a second oscillation axis has to be included, resulting in nodal lines and circles.

A description of the observed oscillations with nodal arrangements is shown in **Fig. 5**. Here, the resonance modes are labelled with two parameters, the number of nodal diameters (**Fig. 5**, 2$^{nd}$ row, first number) and the number of nodal circles (**Fig. 5**, 2$^{nd}$ row, second number).

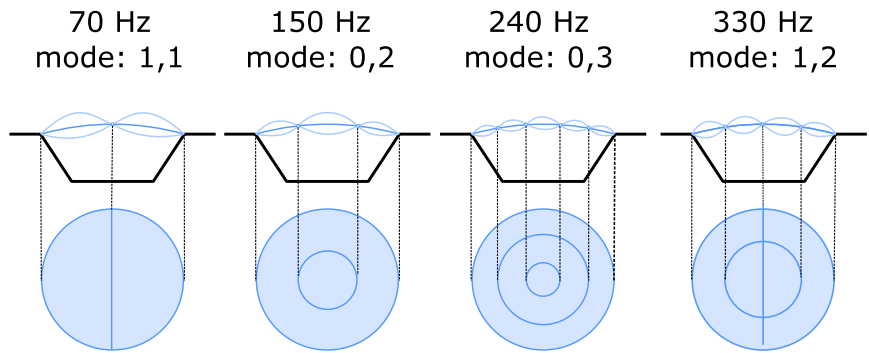

**Figure 5:** Schematic illustration of resonance frequencies (1$^{st}$ row) and resonance modes (2$^{nd}$ row) on the liquid surface in the cwMBR due to vertical oscillation. In the third row, the cwMBR is shown in side view (black line) with the gas/liquid interface in dark blue. Standing waves are shown in lighter blue. The according mode shapes are shown in the 4$^{th}$ row lined with nodal lines and nodal circles, which are not moving while the surrounding liquid is oscillating.

Excitation of the cwMBR at 70 Hz leads to a mode that can be described with one nodal diameter in the center and one nodal circle at the outside, the three-phase contact line, which is therefore numbered (1,1). One side of the liquid surface exhibits an upward movement, the other side a downward movement. This motion can be referred to as rocking [57]. Meinen et al. [64] observed a stationary vertical movement with a similar pattern inside the cwMBR. If excited with $f_1$ = 66 Hz, having a nodal diameter in the reactor center and a nodal circle at the outer ring, good agreement of oscillation mode and the resulting fluid flow can be shown. When excited with 150 Hz, the oscillation pattern fits to the (0,2) mode. In this case, no nodal diameter but two nodal circles, arise, one at the outer edge and one close to the center. Consequently, either the inner circular area or the outer ring is moving upwards, whereas the other one moves down respectively. As soon as the excitation frequency is further increased to 240 Hz, the liquid surface shows a (0,3) mode, showing again no nodal diameter, but one additional nodal circle compared to the previous (0,2) mode. These two modes without nodal diameters are very similar and therefore challenging to distinguish. The mode at 330 Hz is most reasonably the (1,2) mode, with one nodal diameter and two nodal circles. This is the most complex oscillation pattern shown here, where both rings are split in half. Therefore, the outer left ring is synchronized with the right inner ring and the opposite holds true for the other two rings.

At (0,1) mode (not shown), the entire liquid surface would perform an up-down movement with the highest amplitude in the center. This mode is generally not allowed for incompressible fluids in constrained droplets, since it would equate to a volume change, if the contact line is pinned [55]. For this reason, the first calculated resonance, here calculated with 27.4 Hz (**Tab. 1)**, cannot occur. Additionally, exciting modes with nodal circles seem to be more favorable than modes with nodal diameters summing to the same total nodal number. The (2,1) mode (not shown) could only be observed for increasing signal strength at an excitation frequency of 200 Hz, but showing instable oscillation and spilling of the liquid. This is likely resulting from a mobile three-phase contact line of the liquid in the cwMBR for higher excitation amplitudes, which was previously reported by Noblin et al. [58].

In summary, the oscillating liquid surface in the cwMBR can be described with sufficient accuracy using a model of a vibrated sessile droplet. However, the oscillating three-phase contact line reported by Bostwick and Steen [54,57] is less pronounced in the present case. This can probably be explained by the form of the oscillating liquid. The curvature of the droplet examined in this study is lower, compared to the almost spherical caps in most of the sessile droplet studies [54,58,60]. The liquid cap in the cwMBR is significantly lower than a pure sessile droplet with a reasonably large contact angle, resulting in slight deviations of the oscillation modes.

After characterizing the effects of vertical oscillation on the mode patterns excited on the liquid surface, the resulting mass transfer in the liquid phase is to be examined. Through the oscillation of the liquid interface and the involved oscillation of the center of mass, the fluid is set into motion. Hence, the observed surface oscillation is linked to the bulk oscillation [55,64]. This phenomenon is used to mix and homogenize the bulk phase of the cwMBR.

*3.2. Analysis of mixing time in the capillary wave micro-bioreactor*

To analyze the homogenization performance depending on the frequency and excitation strength, the mixing time ($t_M$) was determined using a colorimetric method. By injecting an ink-solution at the reactor bottom combined with image analysis as previously described in Frey et al. [44] the homogenization process was quantified.

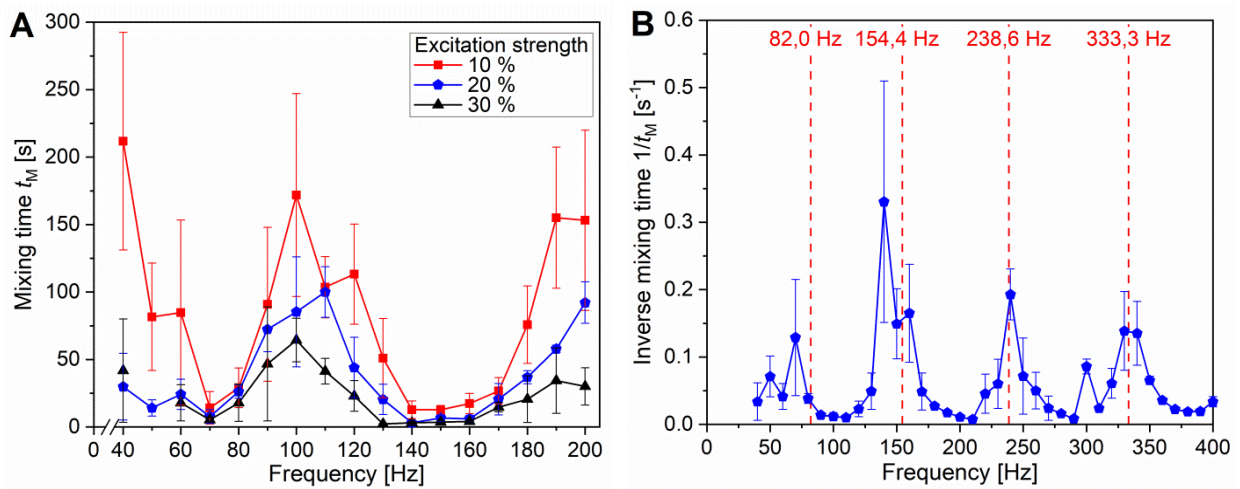

**Figure 6**: Mixing time ($t_M$) depending on different vibration settings for a liquid volume of 8 µL: (A) $t_M$ up to frequencies of 200 Hz with various excitation strengths. (B) Inverse mixing time $1/t_M$ for frequencies up to 400 Hz at excitation strength of 20 %. Calculated resonance frequencies are shown on top in red. All data are given as mean and standard deviation (n = 3).

**Fig. 6A** shows, that $t_M$ strongly depends on the excitation frequency as well as on the excitation strength, with values between 212 s (frequency 40 Hz, excitation strength 10 %) and 2.2 s (130 Hz, 30 %), respectively. Higher excitation strength in general leads to lower mixing times for a given frequency. To avoid leaking of the cwMBR, the excitation strength was only increased up to 30 %. Looking at the frequency, two distinct ranges with low $t_M$, i.e. fast homogenization, are clearly evident: A narrow one around 70 Hz and a broader one around 150 Hz. At these frequencies homogenization is achieved in only a few seconds, while for the remaining frequencies the shown $t_M$ increases.

The zones of rapid homogenization in turn correlate with the theoretical resonance frequencies listed in **Tab. 1**. This characteristic becomes even more apparent by plotting the inverse mixing time $1/t_M$ against the frequency, shown in **Fig. 6B**. Fast homogenization (i.e. larger $1/t_M$) correlates well with the theoretical resonance frequencies marked with dashed lines in red. The actual maxima, however, are at slightly lower frequencies. This observation was already reported for oscillation models, which tend to overestimate the resonance frequency [55,60,63].

To achieve low $t_M$ (i.e. fast mixing), the decisive factor turns out to be the frequency, whereas the excitation strength solely acts as an amplification factor. Only if excited in resonance, effective homogenization can take place. Fastest mixing is achieved at second and third resonance frequencies (140 and 240 Hz), where only nodal circles occur (0,2 mode and 0,3 mode). Here, it has to be taken into account, that the vertical displacement of the vertical vibration system is generally lower for higher frequencies [44]. Hence, the imposed maximum acceleration is lower for higher frequencies.

Additionally, for higher modes with a larger wavenumber, i.e. more nodes, the amplitude of the oscillating liquid surface decreases [58,60].

The oscillation mixing technique enables rapid homogenization within only a few seconds, which is prerequisite to supply microorganisms with substrate avoiding limitation of cell growth and to avoid sedimentation. The applied vertical oscillation mixing can therefore compete with other techniques for small scale mixing reported in the literature. Zhang et al. demonstrated mixing times of less than 30 s for a stirred 150 µL MBR [75]. A 22 µL-reaction chamber was shown to be homogenized within 35 s [76]. The mixing time in a miniaturized stirred tank reactor was reported to be 4.8 s leading to comparable cell growth as in large scale cultivation [77]. A 1000 µL micro-well was mixed within only 1.7 s using orbital shaking [78]. In comparison, typical laboratory shake flasks are reported to be homogenized within one second [79]. Even faster homogenization was only reported for ultrasound or acoustic mixing systems in smaller fluid volumes achieving mixing within milliseconds [45,50,52]. Due to the enhanced energy dissipation, these systems are prone to increase temperature of the liquid causing difficulties for the biological application [50].

After characterizing $t_M$ and its dependency on the frequency and the excitation strength, the influence of fluid viscosity on the mixing is to be evaluated. The viscosity of the cultivation broth can increase during cell cultivation due to high cell densities as well as metabolic products. The mixing performance, in turn, strongly depends on fluid properties, especially the viscosity, influencing the Reynolds number [46] and directly affecting the fluid flow. Therefore, an effective mixing method must be capable of mixing a broad range of fluid viscosities. To investigate the effect of viscosity on the mixing performance, $t_M$ was measured for glycerol solutions of various volume percentages, shown in **Fig. 7**.

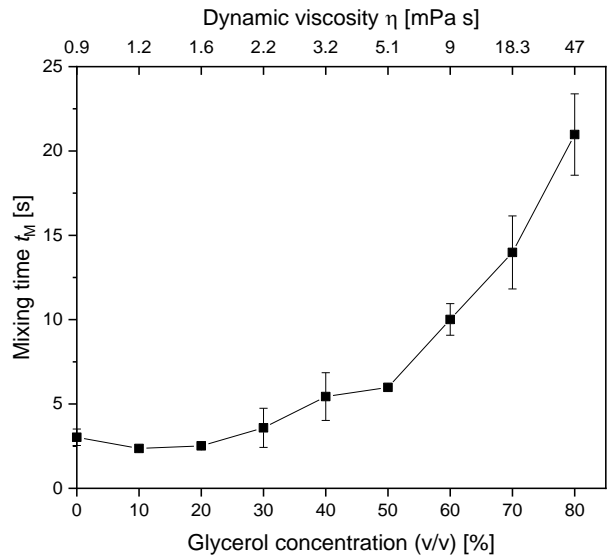

**Figure 7**: Mixing time ($t_M$) depending on the glycerol concentration for an oscillation frequency of 150 Hz, an excitation strength of 20 % and a fluid volume of 8 µL. At these oscillation settings, $t_M$ was found to be minimal, as shown in Fig. 6. $t_M$ in turn was determined using a colorimetric method combined with image analysis as described in [44]. According dynamic viscosities for the investigated glycerol solutions are shown on the upper x-axis, resulting in a non-linear labeling. Values for dynamic viscosities were taken from Segur and Oberstar [80]. Data are shown in triplicates with standard deviation.

Initially, $t_M$ slightly decreases for glycerol concentration of 10 % followed by an equal value of 20 %. For higher glycerol concentrations up to 80 %, $t_M$ increases to about 20 s, with an almost exponential increase between 30 and 80 %. This can be explained by the exponential increase of the dynamic viscosity with the glycerol concentration. This exponential correlation was previously reported by Mugele et al. [81] as well as by Kardous et al. [50], where the particle velocity inside a 400 nL droplet was reduced with increasing glycerol concentrations. The increase in viscosity leads to a damping of the oscillation that likewise affects the mixing time [57,82]. Both, decreased surface oscillation as well as impaired fluid flow, contribute to the rise of $t_M$. Despite the increase for higher glycerol concentration, mixing fluids with a dynamic viscosity of 47 mPa s in 20 s can be regarded as rapid mixing. Cultivations of *E. coli*, in contrast, were reported to have a viscosity of 2 m Pa s after 40 h, stating the end of the exponential growth phase, and up to 6 m Pa s after 100 h [83,84]. Thus, using the oscillation mixing, fluids with sufficiently high viscosities can be homogenized adequately fast to be applicable for a wide range of cultivations with even high cell densities.

*3.3. Oxygen mass transfer analysis*

To support a biologically active environment, substrate limitations inside the cultivation broth have to be avoided. Due to both, the low solubility of oxygen in aqueous culture media and the high oxygen demand in aerobic bioprocesses, ensuring sufficient oxygen supply is often considered as the key challenge for transport processes in bioreactors [41]. To characterize and quantify the oxygen transfer into the liquid phase, most commonly the volumetric mass transfer coefficient $k_La$ is used. Using this single parameter, dissolved oxygen (DO) transport performance can be compared across different cultivation systems and scales. Therefore, $k_La$ values were determined for the cwMBR applying different vibration settings using an optical sensor spot, which was read out from below the reactor via an optical fiber (**Fig. 8**).

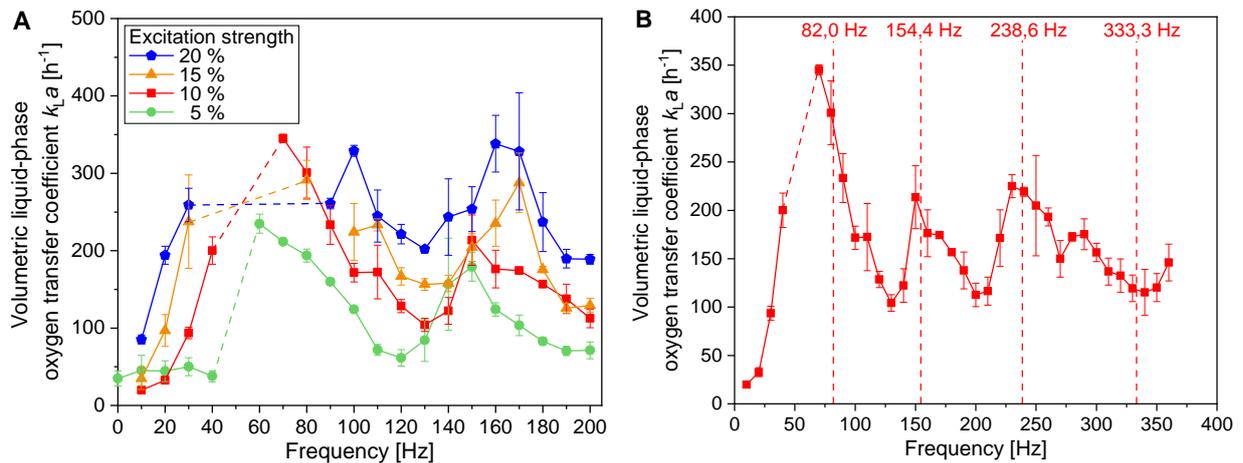

**Figure 8**: Volumetric liquid-phase mass transfer coefficient ($k_La$) in the cwMBR filled with 8 µL for various vibration settings. (A) $k_La$ values for different excitation strengths. Dashed lines indicate areas of increased tendency of liquid spilling, i.e. non reproducible measurements (B) $k_La$ values for an excitation strength of 10 % up to a frequency of 360 Hz with indicated calculated resonance frequencies (Eq. (1)). All data are shown in triplicate with standard deviation.

In **Fig. 8A** the $k_La$ values for different excitation strengths are depicted. Applying the oscillation technique to the cwMBR, $k_La$ values between 19.8 h$^{-1}$ (frequency 10 Hz, excitation strength 10 %) and 345 h$^{-1}$ (70 Hz, 10 %) were measured. For frequencies between 40 to 60 Hz, reproducible measurements could not be performed due to spilling of the liquid (marked by dashed lines). Comparable to $t_M$ shown in **Fig. 6**, the $k_La$ values strongly depend on both the frequency as well as the excitation strength. For a given frequency, increasing the excitation strength enhances the $k_La$. As previously reported, two regions for enhanced mass transfer, i.e. high $k_La$ values, are visible: one around 70 Hz and another one around 150 Hz. These areas of increased $k_La$, again, correlate well with the calculated resonance frequencies. The dependency gets even more apparent, if $k_La$ values are plotted for a wider frequency band at an excitation strength of 10 % shown in **Fig. 8B**.

Here, four distinct peaks of large $k_La$ values can be seen. Except for the peak at highest frequency, all $k_La$ maxima fit the theoretical resonance frequencies (compare **Tab. 1**). The last peak is clearly shifted towards lower frequencies. Oxygen transfer is the highest at 70 Hz, both resonance frequencies at 140 Hz and 230 Hz show comparable results. The enhanced mass transfer via vertical oscillation can be explained by two factors. First, the resonance leads to stronger oscillation of the liquid interphase resulting in faster mixing and avoiding concentration gradients. Oxygen is therefore transported throughout the fluid more rapidly. This has already been explained in chapter 3.2 concerning $t_M$. Second, the oscillation leads to a higher specific gas liquid interphase (term $a$ in $k_La$, which represents the gas liquid interphase related to the filling volume) through the standing wave on the liquid surface, where the actual mass transfer takes place by diffusion. Oxygen transport into the liquid phase is influenced by these two factors, the total specific gas-liquid interphase surface area available for mass transfer as well as the concentration gradient on the gas-liquid interphase. Hence, an enlargement of the liquid surface increases the $k_La$.

In general, oxygen mass transfer can be easily controlled and modified by adjusting the vibrational settings in order to meet the experimental needs and to supply sufficient amounts of oxygen throughout a biological cultivation – without actively gassing the fluid. Compared to other MBR systems reported in literature, the cwMBR shows similar or even higher mass transfer rates. Peterat et al. showed for a 60 µL micro-bubble column reactor $k_La$ values of up to 320 h$^{-1}$ [85]. 96-well microtiter plates were evaluated to have $k_La$ values up to 350 h$^{-1}$ resulting in non-limited growth and protein expression kinetics attained in bacteria and yeast cultivations [86]. Orbitally shaken microtiter plates with specifically designed baffle structures even achieved $k_La$ values up to 460 h$^{-1}$ [87,88]. Applying the vertical vibration technique to a different MBR design entirely constraining the fluid $k_La$ values of more than 1000 h$^{-1}$ were achieved [44]. In the latter case, the fluid was completely surrounded by reactor walls. Hence, stronger excitation could be applied, leading to the increased mass transfer.

In summary, oxygen transfer into the liquid phase of the cwMBR can be tailored via the oscillation settings offering a broad operating window, capable of meeting the demands of biological applications.

*3.4 Tailoring biological growth via oscillation settings in the cwMBR*

The cwMBR in combination with vertical oscillation mixing was shown to achieve mixing times of 2 s and $k_La$ values up to 345 h$^{-1}$ proving its suitability to promote cell growth during a microbial cultivation. To investigate the influence of the oscillation settings alongside with differences in mass transfer and the resulting effects on cell growth, cultivations of *E. coli* BL21 were performed applying varying vibration settings (**Fig. 9**).

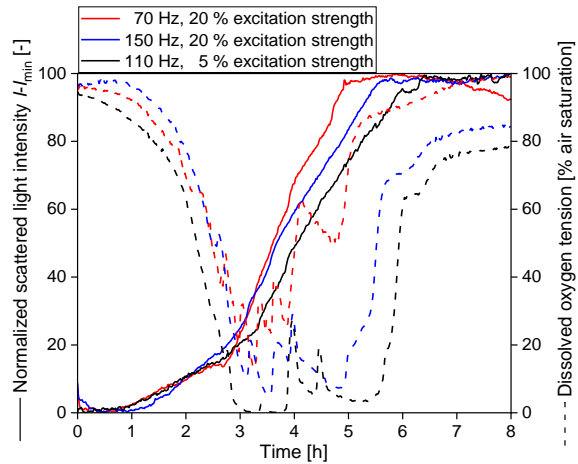

**Figure 9**: Cultivation profiles of *E. coli* BL21 grown in the cwMBR at different oscillation settings (frequency and excitation strength), resulting in conditions with lower mixing performance, medium and highest mass transfer, with regard to chapter 3.2 and 3.3. Solid lines show the scattered light intensities and dashed lines the corresponding dissolved oxygen concentrations. The presented data are mean values of triplicates.

Cultivations were monitored using the oxygen sensor, previously applied for the $k_La$ determination. Additionally, a scattered light measurement was used to monitor the biomass growth over time. To focus on differences in cell growth, the signal was normalized on their maximum value, as previously done for microbial cultivations [44,66].

Comparing the applied oscillation settings for frequency and excitation strength of the cultivations shown in **Fig. 9** with the $k_La$ values determined in chapter 3.3, the corresponding values are 275 h$^{-1}$ for the red curve (frequency 70 Hz, excitation strength 20 %), 183 h$^{-1}$ for the blue curve (150 Hz, 20 %) and 68 h$^{-1}$ for the black curve (110 Hz. 5 %). The effective $k_La$ values tend to be smaller, since the head space of the cwMBR was covered with a silicon barrier to reduce evaporation during the long term cultivations of 8 h. Yet, it is clearly visible, that the course of the cultivations can be influenced by the applied oscillation settings. After a short lag phase at the beginning of the cultivation, the scattered light signals increase after one hour of cultivation showing a parallel course. In the subsequent growth phase, starting after 3 h, the black curve (110 Hz, 5 %) shows the lowest growth rate followed by the blue one (150 Hz, 20 %) and finally the red curve (70 Hz, 20 %). This trend correlates with the applied $k_La$ values: higher $k_La$ values lead to faster cell growth. This statement is supported by the DO concentrations. All three curves start at almost 100 %, corresponding to air saturated liquid. The black curve directly drops negatively exponentially reaching 0 % after 3 h of cultivation. The oxygen signal remains close to zero for another 2.5 h, outlining a clear oxygen limitation. The DO signal for the blue curve initially remains constant for about one hour followed by a decrease to 10 % after around 3 h, just to increase again after 5 h. The red curve decreases, in a similar manner to the black graph, right after the cultivation start, but only reaches 20 % DO. After one hour it increases again with another short decline reaching 100 % at the end of the cultivation.

The repeated declines and increases in the course of DO especially during the exponential growth phase can be explained with metabolic shifts of *E. coli* grown in LB-medium. Due to the low concentrations of sugars, cell metabolism is switched to using amino acids and oligopeptides as carbon source, when sugars are depleted. These are then subsequently depleted sequentially, leading to almost equal changes in DO profiles [89–91].

To conclude, via tuning the oscillation settings, the oxygen limitations are shortened or even prevented (70 Hz, 20 %) resulting in favorable conditions for microbial growth. Additionally, the course of microbial growth was tailored by oscillation settings.

The presented cultivations are in good agreement with previously described data in literature for a 5 µL-MBR [92]. Here, a $k_L a$ value of 60 h$^{-1}$ was applied, which led to a significant oxygen limitation comparable to this study. Cultivations with higher oxygen transport in MBRs show good accordance to the cultivation performed at 70 Hz [93,94].

Even though effects of mass transfer on biological growth could be observed, sensing can be enhanced to more precisely monitor cellular processes. Here, the growth phase so far has only led to a linear increase of the *scattered light* signal which implies limited growth. But observing the exponentially decreasing oxygen concentration, an exponential growth can still be assumed. Hence, the biomass monitoring needs improvement for future setup considerations.

## 4. Conclusions and Outlook

Vertical oscillation effects the formation of characteristic mode patterns on a capillary wave MBR, which strongly depend on the excitation frequency. Using a model of a vertically vibrated droplet, both resonance frequencies and the resonance mode shapes can be described with accuracy. Via adjusting the oscillation settings, particularly the excitation frequency and excitation strength, mixing time and mass transfer can be controlled leading to rapid homogenization and high oxygen transfer rates. Additionally, capillary waves were shown to adjust the cultivation performance of *E. coli* BL21 proving the applicability of the demonstrated cwMBR for biological use.

However, conducting biological processes in an 8 µL droplet causes challenges, particularly in monitoring and sensor integration. Accessibility and sensor read out was facilitated throughout the planar reactor bottom as well as by the high transparency, achieved by manufacturing the cwMBR by fs-LDW. This lead to a significant improvement compared to the previously applied spherical reactor shape [44]. Observation of cell growth, however, must be refined for future setup considerations to enable more precise biomass measurements, although sufficiently descriptive experimental data was obtained. Here, the angle between both optical fibers can be adjusted to improve signal validity.

Alternatively, another measurement technology can be applied, for instance impedance spectroscopy, enabling the differentiation between viable and dead biomass. To enhance reproducibility and enable more precise long-term analysis, an active liquid level control to counterbalance liquid loss due to evaporation can be integrated. The system yields the full benefits, when multiple cwMBRs are operated in parallel. It can then be applied for biopharmaceutical screenings and analysis, such as the testing of active ingredient formulations on cell cultures.

**Declaration of interest**

The authors declare that they have no competing interests.

**Authors' contributions**

LJF and RK conceived the study. LJF and DV conducted the main lab work, the mixing time and mass transfer characterization, the cultivation of *E. coli*, the literature review and the manuscript preparation. SM and AD designed and manufactured the Foturan® glass-based cwMBR. BM and TM implemented the online DO sensor on the cwMBR bottom. DR implemented the use of the cwMBR into the laboratory environment and took the photographs of the different oscillation patterns. RK and JHG supervised the study, participated in its design and coordination and helped to draft and revise the manuscript. All authors read and approved the final manuscript.

**Acknowledgements**

The authors gratefully acknowledge financial support from the German Research Foundation (DFG) within the project *Development of micro-reactors for biopharmaceutical applications* (KR 1897/5-1 (Rainer Krull), DI 1934/9-1 (Andreas Dietzel)). Dr. Rebekka Biedendieck, Institute of Microbiology, Technische Universität Braunschweig, Germany, is additionally thanked for providing the *Escherichia coli* BL21 strain.

**Abbreviations**

| | |
|---|---|
| *cwMBR* | capillary wave micro-bioreactor |
| *DO* | dissolved oxygen |
| *fs* | femtosecond |
| *LDW* | Laser Direct Writing |
| *MBR* | micro-bioreactor |
| *n* | half of the wavelength on the droplet surface |

**Nomenclature**

| | |
|---|---|
| *a* | specific gas liquid interphase ($m^2\ m^{-3}$) |
| $f_n$ | resonance frequency (Hz) |

*g*          gravitational force (m s$^{-2}$)

*h*          height of fluid (m)

$k_L a$       volumetric liquid-phase mass transfer coefficient (h$^{-1}$)

*L*          length of the surface profile (m)

$t_M$        mixing time (s)

*γ*          surface tension between liquid and gas (N m$^{-1}$)

*ρ*          liquid density (kg m$^{-3}$)